 \documentclass[final,5p,times,twocolumn,numbers]{elsarticle}

\usepackage{amssymb}
\usepackage{lipsum}
\usepackage{graphicx}
\usepackage{dcolumn}
\usepackage{bm}
\usepackage{bbm}
\usepackage{physics}
\usepackage{amsfonts}
\usepackage[textsize=\scriptsize]{todonotes}
\usepackage{hyperref}
\usepackage{cleveref}
\usepackage{float}
\usepackage[draft,defaultcolor=black]{changes}

\setcitestyle{numbers,square}

\journal{Results in Physics}

\begin{document}

\begin{frontmatter}



\title{Arbitrary state preparation in quantum harmonic oscillators using neural networks}

\author[label1]{Nicolas Parra-A}
\affiliation[label1]{organization={Grupo de Superconductividad y Nanotecnología, Departamento de Física, Universidad Nacional de Colombia},
            city={Bogotá},
            country={Colombia}}

\author[label2]{Vladimir Vargas-Calderón}
\affiliation[label2]{organization={D-Wave Systems},
            city={Burnaby},
            state={British Columbia},
            country={Canada}}

\author[label1]{Herbert Vinck-Posada}

\begin{abstract}
Preparing quantum states is a fundamental task in various quantum algorithms. In particular, state preparation in quantum harmonic oscillators (HOs) is crucial for the manipulation of qudits and the implementation of high-dimensional algorithms.
  In this work, we develop a general methodology for quantum state preparation in an HO coupled to an auxiliary qubit, guaranteeing that any target state is physically preparable.
  Both the qubit and the HO are driven by two lasers with time-dependent phase modulation.
  The modulation times and phase values are generated by a neural network whose input is the desired target state.
  In contrast to conventional quantum control approaches, this framework eliminates the need for per-instance optimization of the control protocol.
  Instead, the control parameters required to prepare an arbitrary quantum state of the HO are obtained directly from a single forward pass through the neural network.
  Specifically, we present results for preparing arbitrary qubit, qutrit, and qudit ($n=4$) states in the HO, achieving average fidelities of 99.99\%, 99.5\%, and 98.9\%, respectively, across random target states.
\end{abstract}



\begin{keyword}
Quantum state preparation \sep Harmonic oscillators \sep Neural networks, Qudits \sep Jaynes--Cummings model



\end{keyword}

\end{frontmatter}




\section{Introduction}

State preparation refers to the process of transitioning a system from an initial state to a target state through a series of interactions applied to the system.
This process is a fundamental step in many quantum technologies and, in particular, in quantum algorithms~\cite{qstate_preparation_functions}.
Improperly preparing the initial states can significantly affect the algorithm's behavior \cite{build_Quantum_supercomputer}.
For instance, in the simulation of physical systems, an initial state must be prepared before evolving it \cite{Quantum_Simulators,PRXQuantum.4.027001}.
In quantum machine learning algorithms, information needs to be encoded into an initial state, which is then used to make predictions \cite{QML-real-world, vlado_classification_measurments}.
Other examples requiring state preparation include algorithms for solving equations \citep{PhysRevA.110.012430, Krovi2023improvedquantum}, quantum chemistry \citep{quantum_chemistry, qt_chesmestry_2}, quantum metrology \cite{qt_metrology, qt_metrology_n_qubit}, and other applications such as quantum communications and quantum memories \citep{Dooley:17, fast_storage_cavities, qt_communication}.

The literature offers various approaches to determine the physical interactions needed to guide a system from an initial state to a target state.
Some of these works employ Krotov's methods \cite{PhysRevA.90.023824}, quantum steering \cite{qt_preparation_steering}, quantum circuits \citep{PhysRevLett.127.090504, PRXQuantum.2.010101}, reinforcement learning \citep{Porotti2022deepreinforcement, PhysRevResearch.5.043002} and other techniques \citep{PhysRevA.100.023410,PhysRevA.109.022441, PhysRevA.109.042401,stp_matrix_products, zhou2024optimalcontrolopenquantum}.
\added{However, these methods require the construction of an optimization 
problem for each target state--which makes them less suitable for 
applications requiring rapid or repeated state preparation across 
many different targets--or techniques that require considerable 
computational resources.}
Other, more analytical techniques aim to solve the dynamics required to reach a specific target state from an initial state \citep{PhysRevLett.76.1055,PhysRevLett.71.1816}.
These methods often yield a separate family of equations for each target state, making them impractical for multiple applications.
Even though approximate methods have been proposed as alternatives \cite{PhysRevLett.122.020502, Bausch2022fastblackboxquantum,sparse_preparation, PhysRevA.110.032609}, a truly practical solution demands a technique capable of rapidly receiving the target state on demand and predicting the interactions required in the physical system, ensuring efficient state preparation.

In this work, we propose using a neural network in which the input is the target state to be prepared, and the output consists of the interactions required to achieve that state.
We focus on preparing states in $n$-dimensional systems, also called qudits.
These systems have attracted growing interest in recent years due to their enhanced capacity for encoding information, robustness to noise, and applications in high-dimensional quantum computing \citep{high_level_qc,Ringbauer2022,Chen_2021, many_qudits_paper, goswami2025quditbasedscalablequantumalgorithm}.
The challenge of using these systems lies in the complexity of their manipulation, especially when it comes to state preparation \cite{https://doi.org/10.1002/qute.201900038}.
In particular, we consider using harmonic oscillators (HOs) to realize control on a finite subset of an HO's levels.

In general, HOs cannot be fully controllable under a finite set of steps \citep{no_control_infinite_dimensional_systems,PhysRevLett.71.1816}.
To address this, an auxiliary system is employed.
In the case of an HO, an auxiliary qubit is used to interact with the oscillator over a finite period.
Under certain approximations, such as the rotating wave approximation \cite{Shore_2011}, an interacting Jaynes-Cummings-type system is formed, which is known to be \added{approximately controllable within a truncated $n$-level manifold}~\cite{jc_controlabillity}.

Superconducting quantum electrodynamics circuits can realize such systems~\citep{garcía_ripoll_2022} with Rabi oscillation periods of tens of nanoseconds~\citep{blais2021circuit}.
Both the HO and the qubit can be driven by microwave control lines where the phase can be modulated rapidly at a few nanoseconds or at the sub-nanosecond scale~\citep{kalfus2020highfidelity}.
As a result, multiple phase switches can be executed within a single Rabi period.
This substantially simplifies numerical simulations of the system dynamics: between successive switches, the control Hamiltonian reduces to that of a resonant drive with fixed amplitude and phase.
The resulting piecewise time-independent form of the Jaynes-Cummings-type Hamiltonian can be straightforwardly integrated to obtain the system's dynamics.

\begin{figure}
  \includegraphics[width=\columnwidth]{ 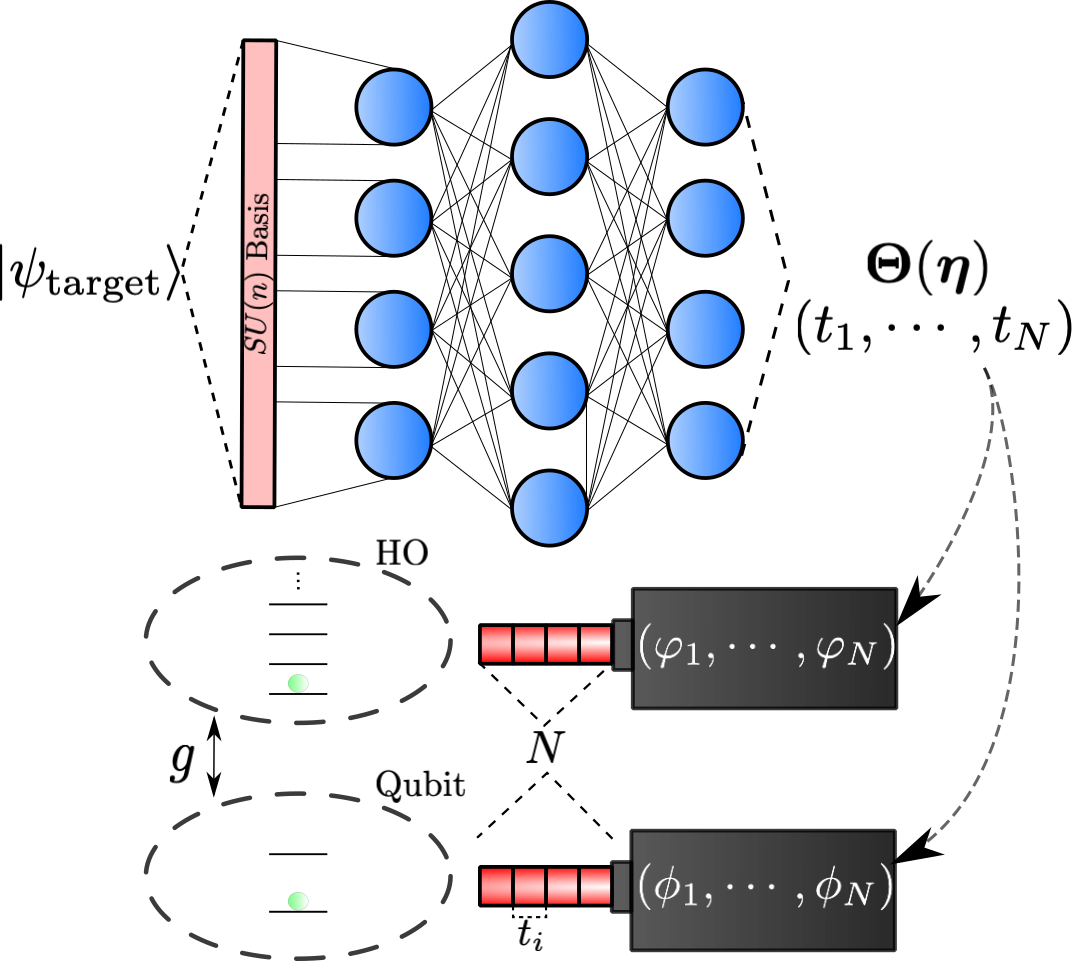}
  \caption{Scheme of the proposed methodology. The target state passes through a layer that computes the expectation values of the elements of the $SU(n)$ basis.
    These values are fed into a neural network with parameters $\vb*{\eta}$. The network outputs the phases of a sequence of $N$ pulses applied to the HO ($\vb*{\varphi}$) and to the ancillary qubit ($\vb*{\phi}$), as well as the switch times $\vb*{t}$ at which phase changes occur.
  These pulses are applied to the system's initial state $\ket{0,0}$, leaving the harmonic oscillator prepared in the target state. More details are given in section \ref{seccion 2}.}
  \label{fig:nn-structure}
\end{figure}

Taking this into account, in our proposal, the neural network will receive the target state to be prepared in the HO as input, and the output will consist of the parameters of the drivings applied both to the HO and to the ancillary qubit to evolve the HO into the target state.
In \cref{fig:nn-structure}, we show a schematic representation of the methodology.

The quality or precision of our model is evaluated using a state fidelity criterion, which is far superior to other neural-network-based proposals in the literature (cf. the noiseless case in \cite{wang2025machinelearningassistedpulsedesignstate}).
Additionally, we investigate how the number of pulses in the sequence enhances the precision with which the state can be  prepared. \added{The term ``arbitrary'' refers to target states within the finite, truncated Hilbert space considered in the model. Its use is justified by the controllability result of Pinna and Panati~\cite{jc_controlabillity}, which shows that the Jaynes--Cummings-type dynamics is approximately controllable in such a truncated $n$-level manifold. Therefore, the neural network is trained to predict, without per-instance optimization, pulse sequences for representative cases ($n=2,3,$ and $4$) where this theoretical reachability applies.}

The structure of this paper is as follows. In \cref{seccion 2}, we describe the physical model, introduce the control parametrization, and detail the full training pipeline used to learn pulse sequences from target states. In \cref{seccion 3}, we then present and analyze the numerical results for qubit, qutrit, and qudit ($n=4$) state preparation, including a study of how the number of pulses affects the achievable fidelity. Finally, in \cref{seccion 4} we summarize the main findings, discuss experimental considerations and limitations, and outline possible directions for future work.

\section{Method}\label{seccion 2}

In this section, we detail our proposal for using a neural network to predict the pulse parameters necessary to bring an HO coupled to a qubit to a separable state such that the HO state is any target state, when the HO basis is truncated to the first $n$ levels.

The system under consideration, i.e., an HO coupled to a qubit, can receive pulses on both the oscillator and the qubit.
The basis of the quantum states of this system is $\ket{\alpha} \otimes \ket{\beta}$, where $\ket{\alpha} \in \{\ket{0},\cdots,\ket{n-1}\}$ and $\ket{\beta} \in \{\ket{0},\ket{1}\}$.
Note that we truncate the HO levels up to $n$ excitations to be able to represent states and operators in a computer as vectors and matrices.
Sometimes we drop the tensor product notation to save space. Modeling the HO-qubit interaction as a Jaynes-Cummings-type system, the Hamiltonian describing this system is \cite{ivan_jc_hamiltonian} ($\hbar = 1$):
\begin{multline}
  \hat{H}(t) = g (\hat{a}\hat{\sigma}_{+} + \hat{a}^{\dagger}\hat{\sigma}_{-}) + \zeta(t) (e^{i\phi(t)} \hat{\sigma}_{-} + e^{-i\phi(t)} \hat{\sigma}_{+}) \\
  + \xi(t)(e^{i\varphi(t)}\hat{a} +  e^{-i\varphi(t)}\hat{a}^{\dagger}) + \Delta_c \hat{n} + \Delta_{\omega_{eg}}\frac{\hat{\sigma}_z}{2},
\end{multline}
where $\hat{a}$ is the HO annihilation operator ($\hat{a}^\dagger$ is the creation operator), $\hat{\sigma}_+$ and $\hat{\sigma}_-$ are the raising and lowering operators of the qubit, $g$ is the oscillator-qubit interaction constant, $\zeta$ and $\xi$ are the amplitudes of the driving pulses, and  $\phi,\varphi$ are the phases of the pulses for the qubit and HO, respectively.
$\Delta_c $ and $\Delta_{\omega_{eg}}$ are the detunings of the HO and the qubit with respect to the driving lasers, respectively.

In this work, we assume resonance of the qubit and the HO with the laser ($\Delta_{\omega_{eg}}=\Delta_c=0$).
Further, we assume constant and equal amplitudes for both the HO and the qubit, i.e., $\zeta(t)=\xi(t) = \frac{\Omega}{2}$, for some pumping constant $\Omega$.
We model the phases of the drivings with a composition of pulses \cite{LEVITT198661}---a technique to increase the robustness of quantum control by relying on multiple constant pulses---using piecewise constant functions of the form $\phi(t) = \phi_i$ for $t_{i-1} < t < t_{i}$, defined by the phase values $\{\phi_i\}_{i=1}^N$ and by the switch times $\{t_i\}_{i=1}^N$, where $t_0=0$.
We assume synchronization of both lasers, so that the switch times are shared for both lasers, but the phase values for the HO are independent parameters $\{\varphi_i\}_{i=1}^N$.
The total evolution time to prepare a target state $\ket{\psi_{\text{target}}}$ in the HO is $t_N$.
As the initial state, we always consider the state $\ket{0,0}$ because this is typically the state of the unexcited system at low temperatures.

With these considerations, the evolution to a target state is given by:
\begin{align}
  \begin{aligned}
    \rho &= \added{\ensuremath{\Tr_\text{q}\left[\ketbra{\psi(\vb*{\Theta})}\right]}}\\
    &\added{\ensuremath{\equiv \Tr_\text{q}\left[\prod_{i=N}^1 e^{i\hat{H}(\phi_i,\varphi_i)(t_{i-1}-t_i)}
    \ketbra{0,0}
    \prod_{i=1}^N e^{-i\hat{H}(\phi_i,\varphi_i)(t_{i-1}-t_i)}\right]}}
  \end{aligned}\label{eq:pulse_application}
\end{align}

where $\Tr_\text{q}[\cdot]$ is the partial trace of the qubit degrees of freedom, $\vb*{\Theta}$ is shorthand notation for the pulse parameters $\{\phi_i\}_{i=1}^N\cup\{\varphi_i\}_{i=1}^N\cup\{t_i\}_{i=1}^N$, and $\hat{H}(\phi,\varphi)$ makes the parameterization of the Hamiltonian explicit. \added{Here, $\ket{\psi(\vb*{\Theta})}$ denotes the final pure state generated by the pulse sequence, $\ket{\psi(\vb*{\Theta})}= \hat{U}(\vb*{\Theta}) \ket{0,0}$.}
Parameters satisfying $\rho = \rho_\text{target} := \ketbra{\psi_\text{target}}$ do exist, as pointed out by Pinna and Panati \citep{jc_controlabillity}, who show that this preparation can be achieved, although, in general, the parameters must be functions of time.
Since we only consider the phase modulation of both drivings, optimal constant pulse parameters might not satisfy \cref{eq:pulse_application}.

To achieve high-fidelity preparation of target states in the HO, we maximize
\begin{align}
  F = \Tr\left[\ketbra{\psi_\text{target}}\Tr_{\text{q}}\left(\ketbra{\psi(\vb*{\Theta})}\right)\right].\label{eq:fidelity_first}
\end{align}

\added{From this point onward, we refer to the infidelity ($1-F$) for convenience in the subsequent analyses.}

There are many state-of-the-art alternatives for determining the pulse sequence required to prepare a state \citep{PhysRevA.90.023824, qt_preparation_steering,PhysRevA.83.032302, PRXQuantum.2.010101,Porotti2022deepreinforcement, PhysRevResearch.5.043002,PhysRevA.100.023410,PhysRevA.109.022441}.
However, these methods rely on iterative techniques that optimize a cost function, whose
dimension increases as larger states are targeted, for each target state.
The fact that each target state requires a separate solving routine makes these algorithms less suitable for real-world implementations requiring rapid encoding and/or state preparation across
many different targets, such as in quantum machine learning~\cite{QML-real-world}.
\added{In contrast, the proposed neural-network approach shifts the computational burden
to a one-time offline training phase, after which control parameters for any new target
state are obtained in a single forward pass. This makes the per-state inference cost
independent of the particular target state, which is advantageous in scenarios requiring
repeated or rapid state preparation across many different targets.}
In this work, we propose using a neural network to predict the pulse parameters
$\vb*{\Theta}$ as a function of the input target state.

Our neural network is, thus, a parameterized function $f_{\vb*{\eta}}:\mathbb{R}^d\to\mathbb{R}^{3N}$, where $d$ real numbers characterize the information in $\ket{\psi_{\text{target}}}$ and $3N$ real numbers specify the Hamiltonian parameters.
$\vb*{\eta}$ are the neural network parameters (for more details, see appendix~\ref{app:nns}).
We can thus write the \added{infidelity} in~\cref{eq:fidelity_first} explicitly as a function of these parameters as
\begin{align}
  1 - F(\psi_\text{target}, \vb*{\eta}) = \expval{\rho_\text{prepared}(\vb*{\eta})}{\psi_\text{target}},\label{eq:fidelity_final}
\end{align}
where $\rho_\text{prepared}(\vb*{\eta})$ is the state prepared in the HO, given by
\begin{align}
  \rho_\text{prepared}(\vb*{\eta}) = \Tr_{\text{q}}\left(\ketbra{\psi(\vb*{\Theta}(\vb*{\eta}))}\right).
  \label{eq:rhoprepared}
\end{align}

The target state is fed into the neural network by passing the expected values of the $d=n^2 -1$ matrices that form a complete basis of $SU(n)$ \cite{lie_algebras_sun} with respect to the target state $\ket{\psi_{\text{target}}}$ (for more details, see appendix~\ref{app:sun}).
A schematic of the information flow through the neural network can be seen in Figure \ref{fig:nn-structure}.
The neural network parameters are optimized using stochastic gradient descent \cite{neural_network_training}.
In the machine learning jargon, carrying out this optimization is called ``training'' the neural network.
Training is done by minimizing the average infidelity over random target states, i.e.,

\begin{align}
  C(\vb*{\eta}) = 1 - \mathbb{E}\left[F(\psi_{\text{target}}, \vb*{\eta}) \right]\label{eq:costfn}
\end{align}
where the expected value in ~\cref{eq:costfn} is taken over random states $\ket{\psi_{\text{target}}}$, sampled from $SU(n)$ according to the Haar measure \cite{haar_measure}.
Thus, we aim to minimize the average infidelity between a set of target states and the states prepared using the phases predicted by the neural network.
As previously mentioned, this is done through stochastic gradient descent, which requires $\grad_{\vb*{\eta}}C(\vb*{\eta})$ to be calculated.
In modern software, this can be easily achieved through automatic differentiation~\citep{jax2018github}.

\section{Results}\label{seccion 3}

In this section, we present results on the performance of the neural network in minimizing the cost function defined in Eq.~(\ref{eq:costfn}). From this point on, we fix the energy scale by choosing $g = 1$.
The training datasets used in the results we present contain 4096 states, sampled using the Haar measure \cite{haar_measure}.
This number was chosen by characterizing how many states are required for the \added{infidelity} to converge, i.e., such that increasing the number of states in the training dataset does not yield any improvement.
Additionally, in our numerical simulations we truncate the HO Hilbert space to a finite dimension $n_\text{comp}>n$ and embed the desired $n$-level target state in this larger computational basis. This accounts for truncation (edge) effects: during the pulse sequence, population can transiently leak into levels above $n$, even though we evaluate the final \added{infidelity} only within the target $n$-dimensional subspace. We set $n_\text{comp}=6$, as previous convergence tests indicated that this cutoff is sufficient for state-preparation simulations with $n=2$, $n=3$, and $n=4$.

\subsection{Number of pulses}\label{subsec:number_pulses}

\begin{figure}
  \centering
  \includegraphics[width=1.0\linewidth]{ 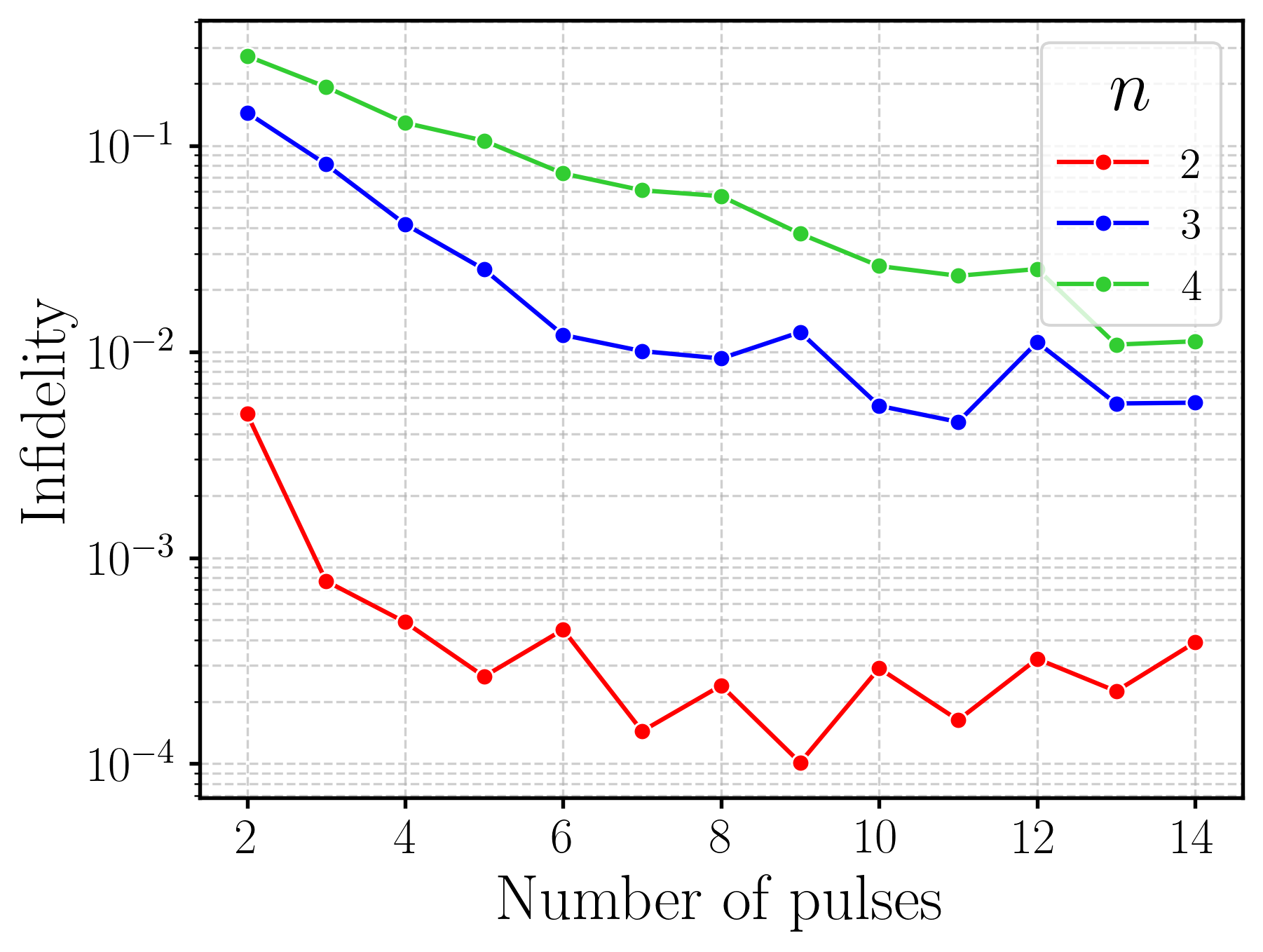}
  \caption{Preparation infidelity using the proposed neural network for 1000 different states as a function of the number of pulses.
  The red line corresponds to preparing qubit states in the HO, the blue line corresponds to preparing qutrit states in the HO, and the green line corresponds to preparing qudit states ($n=4$).}
  \label{fig:infidelity_vs_npulses}
\end{figure}

One of the most common aspects in the literature on pulse composition is that increasing the number of pulses in the sequence improves the quality of the preparation \cite{PhysRevA.100.023410,PhysRevA.109.022441}.
To evaluate this effect, we trained a separate model for each number of pulses ($N=2$--$14$) and qudit dimension ($n=2$--$4$).
For each setting, we sampled 1000 target states from the Haar measure and computed the average \added{infidelity} of the prepared states.
The results are shown in \cref{fig:infidelity_vs_npulses}.

\begin{figure*}[t]
  \centering
  \includegraphics[width=1.02\textwidth]{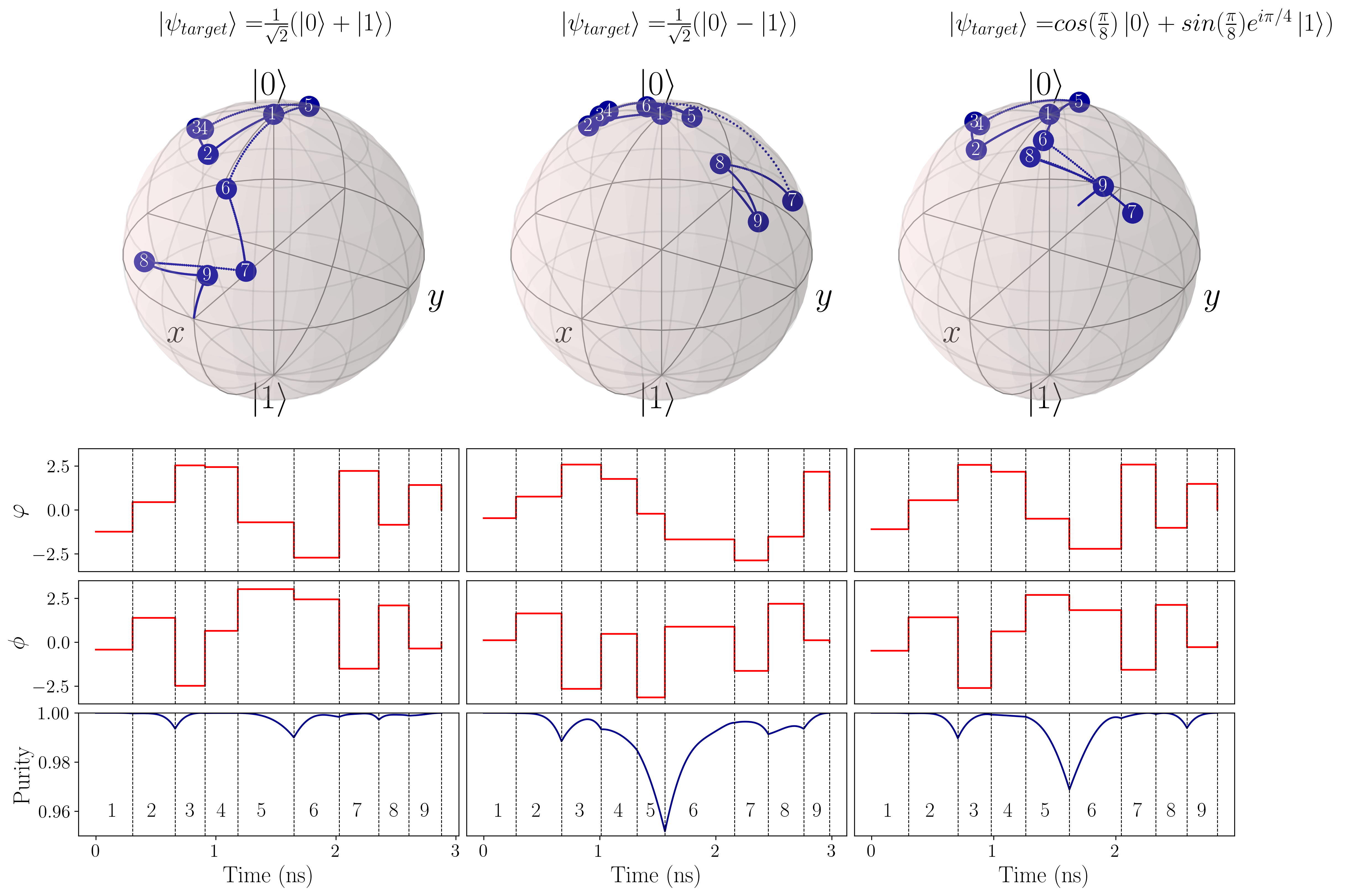}
  \caption{Three representative qubit-state preparation examples (average infidelity $0.01\%$ for $N=9$). The panels show the target state on the Bloch sphere, the neural-network-predicted phase-switching sequences, and the purity evolution during the preparation.}
  \label{fig:qubit_examples}
\end{figure*}

In general, we observe that increasing the number of pulses (or laser phase shifts) \added{reduces the infidelity}, in agreement with the literature. However, for each $n$ there is an optimal point, since increasing the number of pulses beyond that does not further \added{reduce the preparation infidelity}. In the qubit case, the lowest infidelity occurs at $N=9$, achieving an average preparation \added{infidelity} of $0.01\%$. This \added{infidelity} is highly competitive compared to different state-of-the-art techniques. For qutrits, the infidelity already starts to increase noticeably, reaching $0.5\%$ with a sequence of $N=11$, although it remains a low preparation \added{infidelity}. Finally, for qudits with $n=4$, the \added{infidelity} increases slightly further, reaching $1.1\%$ at $N=13$. Although this result may not appear as strong, there are not many techniques that report performance for qudit ($n=4$) preparation, so it provides an interesting starting baseline.
On the other hand, the algorithm appears to exhibit a strong dependence on the number of qudit levels, which is expected because the Hilbert space increases in size; however, given the relatively small increase from $n=3$ to $n=4$, the drop does not seem abrupt. This will require future scaling analyses.

\subsection{Qubit performance}
\label{subsec:qubit_performance}

We now analyze the state-preparation performance in the qubit case ($n=2$). As discussed in \cref{subsec:number_pulses}, the best-performing model is obtained with a sequence of $N=9$ pulses, reaching an average preparation \added{infidelity} of $0.01\%$ over 1000 Haar-random target states.

To better understand how the model works to prepare a state,~\cref{fig:qubit_examples} illustrates three representative target states---the Hadamard states $\ket{+}$ and $\ket{-}$, and the magic state $\ket{T}$---as they are prepared using the pulse sequence predicted by the neural network. The first set of panels shows the trajectory of the qubit prepared state on the Bloch sphere after each pulse.
Notably, individual pulses do not necessarily move the state monotonically closer to the target; rather, the early pulses can drive the system through intermediate directions in state space, while the final pulses perform the fine corrections that align the trajectory with the desired target state.

The figure also reports the phase-switching pattern predicted for each pulse applied to the HO and to the ancillary qubit. No simple, easily recognizable pattern is expected in these phases: they emerge from the learned control strategy and are precisely the degrees of freedom that allow the protocol to compensate accumulated phase and population errors along the sequence.

Finally, the qubit purity is shown throughout the preparation. It typically decreases at intermediate steps, indicating transient entanglement between the qubit and the HO during the control process, and then returns close to unity at the end of the sequence, consistent with the intended final separable state. Achieving such \added{infidelity} preparation for states that are central to quantum information processing---particularly non-stabilizer resource states such as the magic state $\ket{T}$, which are often challenging for alternative preparation strategies---highlights the competitiveness of the proposed approach.

\subsection{Qutrit performance}
\label{subsec:qutrit_performance}

We now analyze the state-preparation performance in the qutrit case ($n=3$). As discussed in \cref{subsec:number_pulses}, the best-performing model is obtained with a sequence of $N=11$ pulses, reaching an average preparation \added{infidelity} of $0.5\%$ over 1000 Haar-random target states. In this setting, the dynamics cannot be visualized as directly as in the qubit case using a Bloch-sphere representation; instead, we characterize the preparation quality through the HO Wigner function,

\begin{equation}
  W(x,p)=\frac{1}{\pi\hbar}\int_{-\infty}^{\infty}\mel{x-y}{\hat{\rho}}{x+y}\,e^{2 i p y/\hbar}\,dy.
\end{equation}

\begin{figure}[H]
  \centering
  \includegraphics[width=0.95\linewidth]{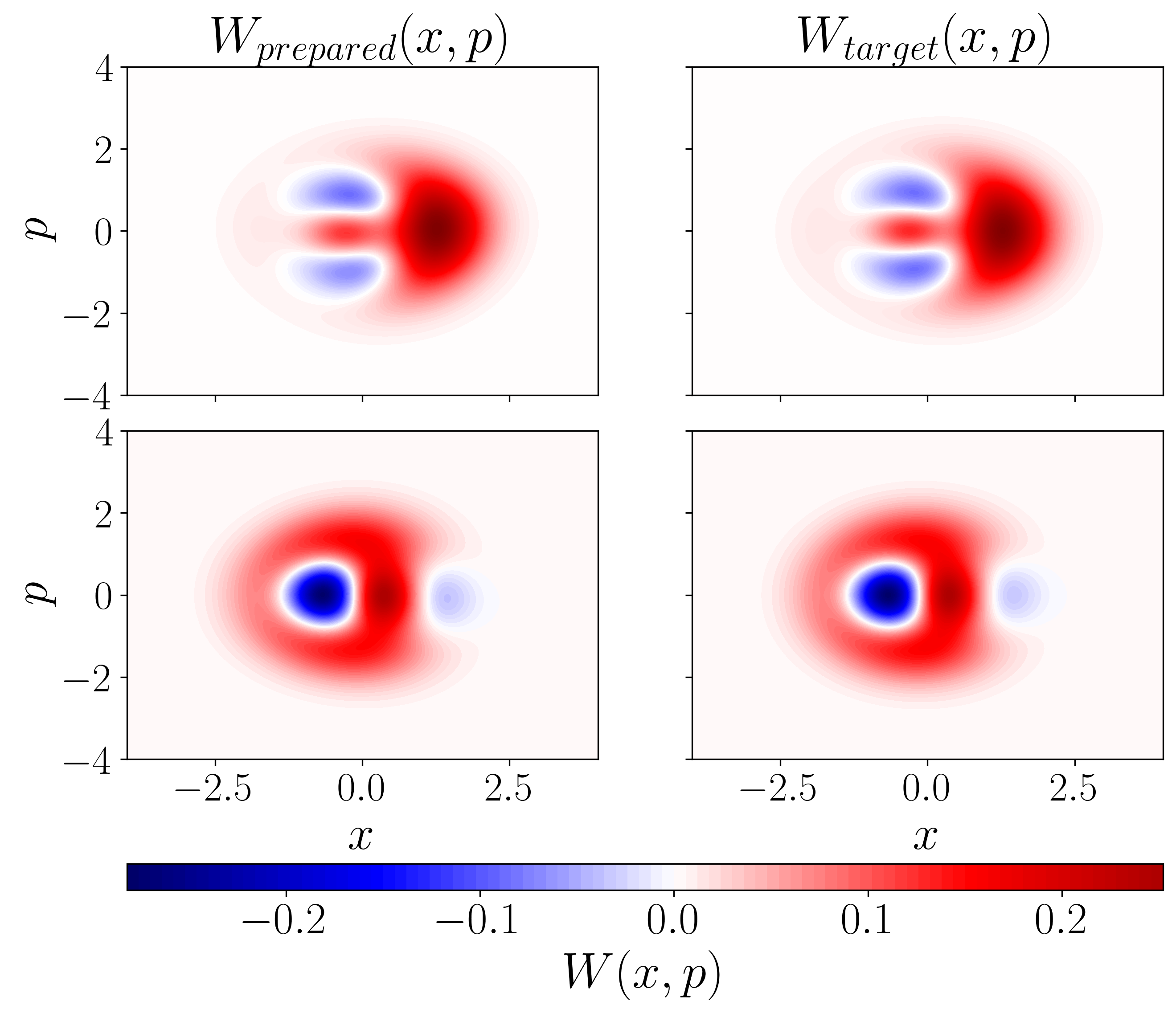}
  \caption{Representative qutrit-state preparation for the target states $\frac{1}{\sqrt{3}} (\ket{0}+\ket{1}-\ket{2})$ and $\frac{1}{\sqrt{3}} (\ket{0}+\ket{1}+\ket{2})$ with average \added{infidelity $0.5\%$} for a pulse sequence of $N=11$. The first column shows the Wigner function of the prepared HO state and the second column shows the target-state Wigner functions.}
  \label{fig:qutrit_examples}
\end{figure}

\Cref{fig:qutrit_examples} compares the HO Wigner function of the prepared state (first column) against that of the target state (second column) for two representative qutrit targets. Each row corresponds to a different target state, $\frac{1}{\sqrt{3}} (\ket{0}+\ket{1}+\ket{2})$ and $\frac{1}{\sqrt{3}} (\ket{0}+\ket{1}-\ket{2})$, respectively. We evaluated the absolute pointwise difference $|W_{\mathrm{target}}(x,p)-W_{\mathrm{prepared}}(x,p)|$, but the differences are practically negligible.

Overall, the prepared and target Wigner functions show excellent agreement: the model reproduces the dominant phase-space support as well as the finer nonclassical structure. The remaining discrepancies are small and typically localized, consistent with the slight \added{increase in average infidelity} compared to the qubit case. These results confirm that the proposed approach remains highly effective for qutrit state preparation.

\subsection{Qudit ($n=4$) performance}
\label{subsec:qudit_performance}

We now turn to the qudit case with $n=4$. As discussed in \cref{subsec:number_pulses}, the best-performing model is obtained with $N=13$ pulses, achieving an average preparation \added{infidelity} of $1.1\%$ over 1000 Haar-random target states. \added{Before discussing these examples, we clarify the sense in which two-qubit entangled states are encoded in the oscillator. The four-dimensional subspace spanned by the first oscillator levels is isomorphic to two logical qubits, and we use the identification $\ket{0}\equiv\ket{00}$, $\ket{1}\equiv\ket{01}$, $\ket{2}\equiv\ket{10}$, and $\ket{3}\equiv\ket{11}$. Under this tensor-product structure, entanglement is defined between the two logical qubits encoded in the oscillator subspace, as in standard bosonic-qubit encodings~\cite{garcía_ripoll_2022}.}

\begin{figure}[H]
  \centering
  \includegraphics[width=1.01\linewidth]{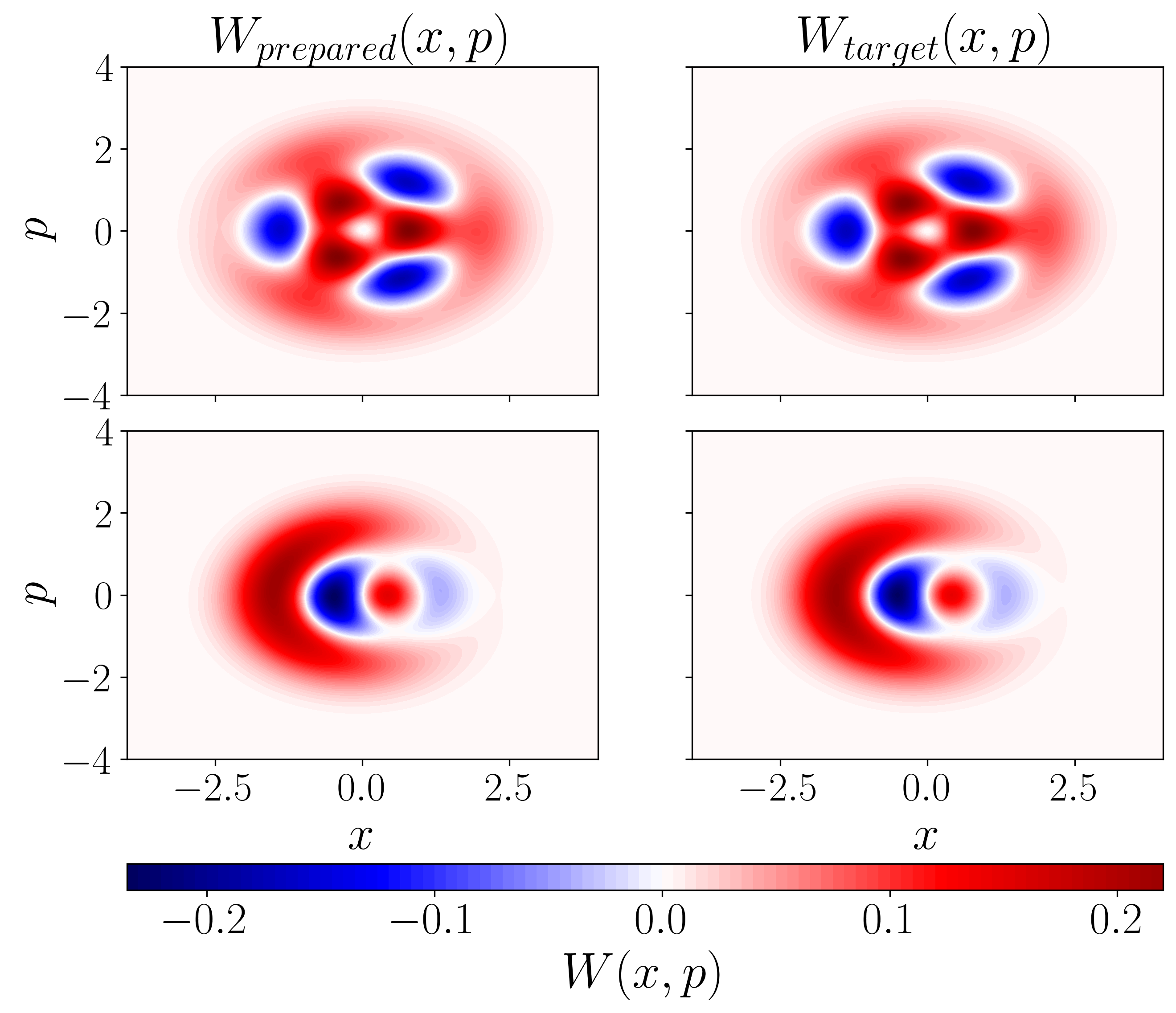}
  \caption{Representative qudit-state ($n=4$) preparation for the target Bell states $\ket{\Phi^{+}}$ and $\ket{\Psi^{-}}$ with average \added{infidelity $1.1\%$} for a pulse sequence of $N=13$. The first column shows the Wigner function of the prepared HO state and the second column shows the target-state Wigner function.}
  \label{fig:qudit4_examples}
\end{figure}

Figure~\ref{fig:qudit4_examples} summarizes representative $n=4$ preparation examples using the same Wigner-function diagnostics as in the qutrit case. \added{With the encoding above, the oscillator states $\ket{\Phi^{+}}=\frac{1}{\sqrt{2}}(\ket{0}+\ket{3})$ and $\ket{\Psi^{-}}=\frac{1}{\sqrt{2}}(\ket{1}-\ket{2})$ correspond to the Bell states $\frac{1}{\sqrt{2}}(\ket{00}+\ket{11})$ and $\frac{1}{\sqrt{2}}(\ket{01}-\ket{10})$, respectively.} Each row corresponds to one of these target states.
The first column shows the Wigner function of the prepared HO state, the second column shows the target-state Wigner function.
Even at $n=4$, the learned pulse sequences are able to capture the main phase-space structure of the target states, including interference patterns and negative regions that signal nonclassicality.
The remaining infidelity is associated with small, spatially localized deviations rather than a global distortion of the Wigner distribution, suggesting that the protocol is already close to the target and primarily limited by fine feature resolution as the Hilbert-space dimension increases.
Nevertheless, achieving this level of performance for $n=4$ qudits is particularly relevant because it demonstrates the ability to prepare nontrivial two-qubit entangled states encoded within the oscillator subspace, a key resource for quantum-information processing. \added{Bell states are especially useful benchmarks in this setting because they are simple enough to interpret under the two-logical-qubit mapping, while still requiring coherent superpositions between separated oscillator levels and well-defined relative phases. Their preparation therefore tests not only population transfer within the four-level manifold, but also the phase coherence needed to generate entanglement-like resources in a single bosonic mode.}

\section{Discussion}
\label{seccion 4}



\subsection{Computational cost and practical advantages}

\begin{table*}[t]
  \footnotesize
  \setlength{\tabcolsep}{4pt}
  \begin{tabular}{p{0.18\textwidth}p{0.20\textwidth}p{0.22\textwidth}p{0.18\textwidth}p{0.13\textwidth}}
    \hline
    \textbf{Method} & \textbf{Offline cost} & \textbf{Online cost per target} 
    & \textbf{Physical control time} & \textbf{Typical fidelity} \\
    \hline
    Optimal control (Krotov/GRAPE-type)~\cite{PhysRevA.90.023824,PhysRevA.83.032302} 
    & None, apart from method setup 
    & Separate iterative optimization per target; $10^2$--$10^3$ iterations, 
    each requiring a full dynamics simulation; total runtime of order minutes 
    to hours per target state
    & Platform-dependent; tens to hundreds of ns in circuit-QED 
    implementations~\cite{PhysRevA.90.023824,blais2021circuit}
    & Infidelities $\leq 10^{-4}$ in benchmark settings \\
    Reinforcement-learning control~\cite{Porotti2022deepreinforcement,PhysRevResearch.5.043002} 
    & Training of a policy or value function; hours of computation 
    & Policy evaluation after training; no full optimal-control loop; 
    sub-second to second-scale inference
    & Platform-dependent
    & Task-dependent; high-fidelity benchmarks reported \\
    This work 
    & One-time neural-network training for each $n$ using 4096 Haar-random states 
    & Single forward pass; $\sim$$100~\mathrm{ms}$ per target state on a standard CPU
    & $N=9$--$13$ phase segments; $t_N$ spans a few Rabi periods, corresponding 
    to $\sim$$10$--$500~\mathrm{ns}$ in circuit-QED with 
    $g/2\pi\sim10$--$100~\mathrm{MHz}$~\cite{blais2021circuit}
    & $99.99\%$ ($n=2$), $99.5\%$ ($n=3$), $98.9\%$ ($n=4$) \\
    \hline
  \end{tabular}
  \caption{\added{Comparison of the relevant time scales for the present approach and 
  representative control strategies. The table separates the offline optimization or 
  training cost, the online cost of producing controls for a new target state, and the 
  physical duration of the applied pulse sequence. The physical control time is hardware 
  dependent for all methods; for the present phase-switching protocol it is set by the 
  number of pulse segments and the experimentally available Rabi scale.}}
  \label{tab:time_scale_comparison}
\end{table*}

\added{A meaningful assessment of the practical value of the proposed method requires
distinguishing three time scales that are often conflated, and which we summarize in
Table~\ref{tab:time_scale_comparison}. The first is the \textit{offline training cost}:
a one-time optimization of the neural-network parameters $\vb*{\eta}$ for a fixed
Hilbert-space dimension $n$, using the training set of 4096 Haar-random states described
above. This cost is incurred once per dimension $n$ and is fully amortized over all
subsequent uses of the trained model. It should be noted that this training cost is
incurred separately for each Hilbert-space dimension $n$, as the network architecture
and training set are adapted to each case; this represents an overhead that is absent in
per-instance optimal-control methods. However, once a model is trained for a given $n$,
it serves as a universal controller for \emph{all} target states in that dimension, so
the training cost is amortized over the full set of subsequent inference calls, in
contrast to GRAPE- or Krotov-type methods, where the dominant computational cost is
incurred at inference time and repeated independently for every new target state. The
practical advantage of the proposed approach therefore depends on the number of distinct
target states required: for applications involving many different target states, the
amortized cost per state of the neural-network approach is substantially lower. The
second is the \textit{online inference time}: once training is complete, the platform
operates in an on-demand regime, where obtaining the pulse parameters $\vb*{\Theta}$ for
a new target state requires only a single forward pass through the network. In our
implementation, this on-demand generation of controls takes approximately
$100~\mathrm{ms}$ on a standard CPU. This stands in sharp contrast with iterative
optimal-control methods such as Krotov or GRAPE, where each new target state requires a
separate optimization loop that typically demands $10^2$--$10^3$ iterations, each
involving a full simulation of the system dynamics, amounting to runtimes of the order
of minutes to hours per target state depending on the system dimension and convergence
criterion~\cite{PhysRevA.90.023824,PhysRevA.83.032302}. The third is the
\textit{physical control time} $t_N$: the total duration of the pulse sequence applied
to the hardware. In our dimensionless units ($g = 1$), a single Rabi period is
$T_\mathrm{Rabi} = 2\pi/g \approx 6.28$, and the full $N=9$--$13$ pulse sequences span
a total time $t_N$ of a few Rabi periods. Mapping to circuit-QED hardware, where
$g/2\pi \sim 10$--$100~\mathrm{MHz}$~\cite{blais2021circuit}, this corresponds to
physical control times of order $10$--$500~\mathrm{ns}$, which is comparable to pulse
durations reported in conventional optimal-control implementations on similar
platforms~\cite{PhysRevA.90.023824}. Thus, the primary advantage of the proposed method
is not a reduction in physical pulse duration, which remains platform dependent and
broadly comparable across approaches, but rather the ability to generate control
parameters on demand for an arbitrary new target state in $\sim$$100~\mathrm{ms}$. This
on-demand capability makes the method particularly suitable for applications requiring
rapid or repeated state preparation, such as quantum machine-learning workflows where the
target state changes with each input datum~\cite{QML-real-world}.}

\subsection{Limitations and outlook}

\added{Although the proposed approach eliminates per-instance optimization, its 
scalability is constrained by two distinct factors that are worth distinguishing. The 
first is the capacity of the neural network itself: the observed increas in average 
infidelity as $n$ increases from 2 to 4 reflects the growing complexity of the mapping 
from target state to pulse parameters, which requires a more expressive model to be 
approximated accurately. This is not a fundamental limitation of the framework. By the 
universal approximation theorem~\cite{universalapprox}, a feedforward neural network 
with sufficient width or depth can approximate any continuous function on a compact 
domain to arbitrary precision. In practice, this means that increasing the network size 
and complexity---larger hidden layers, additional depth, or more expressive 
architectures---can in principle recover the fidelity lost as $n$ grows. Indeed, the 
network architectures used here were already found to grow with $n$ (hidden-layer widths 
of \mbox{[209, 173, 222, 347]} for $n=2$, \mbox{[197, 264, 304]} for $n=3$, and 
\mbox{[321, 442, 425]} for $n=4$; see Appendix~A), and systematic exploration of larger 
architectures at fixed $n$ is a natural next step. The number of pulses $N$ also plays 
a role analogous to network capacity: as shown in Fig.~2, increasing $N$ consistently 
reduces the infidelity, and larger $n$ requires more pulses to reach near-optimal 
performance ($N=9, 11, 13$ for $n=2, 3, 4$ respectively), suggesting that both network 
size and pulse number should be scaled jointly as $n$ increases.}

\added{The second, and more fundamental, constraint is the cost of generating training 
data. Each training sample requires simulating the full quantum dynamics of the HO-qubit 
system in a Hilbert space of dimension $2n_\mathrm{comp}$, where the computational 
cutoff $n_\mathrm{comp} > n$ must be enlarged as $n$ grows to avoid truncation 
artifacts. The cost of each simulation scales as $\mathcal{O}(n_\mathrm{comp}^3)$ per 
pulse segment due to matrix exponentiation, and the dataset size required for 
convergence is also expected to grow with $n$. For the cases studied here, 
$n_\mathrm{comp} = 6$ and 4096 training states were sufficient; however, both 
quantities will need to increase for larger $n$, making classical simulation the 
practical bottleneck rather than the neural network itself. Extending the method to 
the larger subspace dimensions required by bosonic-encoding applications---such as cat 
qubits, GKP codes, or binomial codes, which typically involve tens to hundreds of 
oscillator levels---will likely require strategies beyond direct simulation, such as 
tensor-network-based state evolution, hardware-in-the-loop training, or transfer 
learning across dimensions. These directions are left for future work.}

A central experimental requirement of the proposed protocol is the ability to implement 
rapid and accurate phase updates of the driving fields. Our control Hamiltonian is 
modeled as a sequence of segments with constant amplitude and constant phase; thus, the 
switching time (and the bandwidth of the control chain) sets the shortest achievable 
segment duration and, consequently, the finest time resolution available to the 
controller.

This assumption is well aligned with circuit-QED implementations, where microwave 
sources combined with IQ mixers and modern arbitrary-waveform generators can realize 
phase switches on the nanosecond scale, enabling multiple phase switches within a single 
Rabi period. In this regime, the piecewise-constant model is realistic and the learned 
sequences can be deployed after standard calibration steps (e.g., correcting for IQ 
imbalance~\citep{jolin2020calibration,wu2024situ}, finite rise 
times~\cite{oh2002errors}, and static phase offsets~\citep{gely2024situ}).

However, the technique becomes unsuitable in platforms where the required control 
timescales are substantially shorter than the achievable electronic phase-update time. 
Such is the case of excitonic dynamics in semiconductor quantum 
dots~\citep{zhang2022ultrafast} or polaritonic systems driven by ultrafast 
lasers~\citep{jia2025femtosecond}. In such settings, a digitally switched, 
nanosecond-resolved phase protocol cannot resolve the dynamics, and the 
piecewise-constant approximation breaks down.

Beyond switching speed, additional non-idealities may reduce the achievable fidelity 
in an experiment: (i) decoherence of both the HO and the ancilla during the total 
preparation time $T$; (ii) amplitude and phase distortions due to finite bandwidth and 
filtering; (iii) imperfect measurement of Hamiltonian coupling strengths and detunings; 
and (iv) leakage outside the truncated HO subspace (especially as $n$ increases) and 
residual contributions from terms neglected under the rotating-wave approximation. 
These effects suggest that the reported fidelities should be interpreted as an upper 
bound for a given hardware platform, and that incorporating realistic noise models and 
robust-training strategies will be important for future work.

A more realistic treatment that includes open-system dynamics and imperfect 
control---e.g., decoherence during the pulse sequence as well as errors and jitter in 
the target phases and switch times---will be the subject of future work.

\section{Conclusions}

\added{In this work, we introduced a neural-network-assisted strategy for arbitrary state preparation in a quantum harmonic oscillator (HO) coupled to an ancillary qubit. The key idea is to train a single feed-forward model that receives a desired target state as input and outputs, in a single forward pass, the parameters of a piecewise-constant phase-modulated driving sequence acting on both the HO and the qubit. By building on the controllability of the effective Jaynes--Cummings-type dynamics within a truncated $n$-level manifold, the approach guarantees that the predicted control protocol corresponds to a physically preparable target within the adopted model and truncation.}

\added{Our numerical results show that the method achieves high-quality preparation across Haar-random targets without per-instance optimization. For qubit targets ($n=2$), we obtained an average fidelity of $99.99\%$ (infidelity $0.01\%$) using $N=9$ pulses; for qutrit targets ($n=3$), we reached $99.5\%$ (infidelity of $0.5\%$) with $N=11$ pulses; and for qudit targets with $n=4$, we achieved $98.9\%$ (infidelity $1.1\%$) with $N=13$ pulses. These results, together with the $n=4$ Bell-state examples, indicate that the method can serve as an on-demand primitive for single-qudit preparation and for generating logical entanglement resources within a bosonic mode. Future work will focus on incorporating open-system dynamics, hardware distortions, and robust-training strategies to assess performance under realistic experimental noise.}

\section*{ACKNOWLEDGMENTS}

N. P.-A. and V. V.-C. would like to acknowledge support from the project ``Aprendizaje de Máquina para Sistemas Cuánticos" HERMES code 57792 and QUIPU code 201010040147, and H. V.-P. from the project ``Ampliación del uso de la mecánica cuántica desde el punto de vista experimental y su relación con la teoría, generando desarrollos útiles para metrología y computación cuántica a nivel nacional" BPIN code 2022000100133. We sincerely thank D. Martínez-Tibaduiza and D. B. Anghelo-Rodriguez for their valuable discussions, comments, and suggestions on this work.

\bibliographystyle{elsarticle-num}
\bibliography{apssamp}

@misc{qstate_preparation_functions,
      title={Quantum state preparation for multivariate functions}, 
      author={Matthias Rosenkranz and Eric Brunner and Gabriel Marin-Sanchez and Nathan Fitzpatrick and Silas Dilkes and Yao Tang and Yuta Kikuchi and Marcello Benedetti},
      year={2024},
      eprint={2405.21058},
      archivePrefix={arXiv},
      primaryClass={quant-ph},
      url={https://arxiv.org/abs/2405.21058}, 
}

@article{wu2024situ,
  title={In situ mixer calibration for superconducting quantum circuits},
  author={Wu, Nan and Lin, Jing and Xie, Changrong and Guo, Zechen and Huang, Wenhui and Zhang, Libo and Zhou, Yuxuan and Sun, Xuandong and Zhang, Jiawei and Guo, Weijie and others},
  journal={Applied Physics Letters},
  volume={125},
  number={20},
  year={2024},
  publisher={AIP Publishing}
}

@article{jolin2020calibration,
	author = {Jolin, S. W. and Borgani, R. and Thol{\'e}n, M. O. and Forchheimer, D. and Haviland, D. B.},
	doi = {10.1063/5.0025836},
	eprint = {https://pubs.aip.org/aip/rsi/article-pdf/doi/10.1063/5.0025836/14886296/124707\_1\_online.pdf},
	issn = {0034-6748},
	journal = {Review of Scientific Instruments},
	month = {12},
	number = {12},
	pages = {124707},
	title = {Calibration of mixer amplitude and phase imbalance in superconducting circuits},
	url = {https://doi.org/10.1063/5.0025836},
	volume = {91},
	year = {2020}}

@article{gely2024situ,
  title = {In situ characterization of qubit-drive phase distortions},
  author = {Gely, M.F. and Litarowicz, J.M.A. and Leu, A.D. and Lucas, D.M.},
  journal = {Phys. Rev. Appl.},
  volume = {22},
  issue = {2},
  pages = {024001},
  numpages = {6},
  year = {2024},
  month = {Aug},
  publisher = {American Physical Society},
  doi = {10.1103/PhysRevApplied.22.024001},
  url = {https://link.aps.org/doi/10.1103/PhysRevApplied.22.024001}
}

@article{jia2025femtosecond,
	author = {Jia, Haoyuan and Cao, Junhui and Chen, Fei and Peng, Fangying and Li, Yihui and Xu, Yihan and Chen, Leizhu and Ye, Ziyu and Zhao, Xianyan and Zhang, Shian and Jing, Jietai and Xu, Hongxing and Chen, Zhanghai and Byrnes, Tim and Li, Hui and Kavokin, Alexey and Wu, Jian},
	doi = {10.1093/nsr/nwaf493},
	eprint = {https://academic.oup.com/nsr/article-pdf/13/1/nwaf493/65388832/nwaf493.pdf},
	issn = {2095-5138},
	journal = {National Science Review},
	month = {11},
	number = {1},
	pages = {nwaf493},
	title = {Femtosecond coherence dynamics of exciton--polaritons},
	url = {https://doi.org/10.1093/nsr/nwaf493},
	volume = {13},
	year = {2025}}

@article{zhang2022ultrafast,
  title={Ultrafast exciton transport at early times in quantum dot solids},
  author={Zhang, Zhilong and Sung, Jooyoung and Toolan, Daniel TW and Han, Sanyang and Pandya, Raj and Weir, Michael P and Xiao, James and Dowland, Simon and Liu, Mengxia and Ryan, Anthony J and others},
  journal={Nature materials},
  volume={21},
  number={5},
  pages={533--539},
  year={2022},
  publisher={Nature Publishing Group UK London}
}

@article{oh2002errors,
  title = {Errors due to finite rise and fall times of pulses in superconducting charge qubits},
  author = {Oh, Sangchul},
  journal = {Phys. Rev. B},
  volume = {65},
  issue = {14},
  pages = {144526},
  numpages = {5},
  year = {2002},
  month = {Apr},
  publisher = {American Physical Society},
  doi = {10.1103/PhysRevB.65.144526},
  url = {https://link.aps.org/doi/10.1103/PhysRevB.65.144526}
}

@article{qt_metrology_n_qubit,
  title = {Quantum metrology with an $N$-qubit $W$ superposition state under noninteracting and interacting operations},
  author = {Li, Yan and Ren, Zhihong},
  journal = {Phys. Rev. A},
  volume = {107},
  issue = {1},
  pages = {012403},
  numpages = {9},
  year = {2023},
  month = {Jan},
  publisher = {American Physical Society},
  doi = {10.1103/PhysRevA.107.012403},
  url = {https://link.aps.org/doi/10.1103/PhysRevA.107.012403}
}

@inproceedings{Dooley:17,
	author = {Shane Dooley and William J. Munro and Kae Nemoto},
	booktitle = {Quantum Information and Measurement (QIM) 2017},
	doi = {10.1364/QIM.2017.QW3A.5},
	journal = {Quantum Information and Measurement (QIM) 2017},
	pages = {QW3A.5},
	publisher = {Optica Publishing Group},
	title = {Quantum metrology including state preparation and readout times},
	url = {https://opg.optica.org/abstract.cfm?URI=QIM-2017-QW3A.5},
	year = {2017}}

@ARTICLE{kalfus2020highfidelity,
  author={Kalfus, William D. and Lee, Diana F. and Ribeill, Guilhem J. and Fallek, Spencer D. and Wagner, Andrew and Donovan, Brian and Ristè, Diego and Ohki, Thomas A.},
  journal={IEEE Transactions on Quantum Engineering}, 
  title={High-Fidelity Control of Superconducting Qubits Using Direct Microwave Synthesis in Higher Nyquist Zones}, 
  year={2020},
  volume={1},
  number={},
  pages={1-12},
  doi={10.1109/TQE.2020.3042895}}

@article{blais2021circuit,
  title = {Circuit quantum electrodynamics},
  author = {Blais, Alexandre and Grimsmo, Arne L. and Girvin, S. M. and Wallraff, Andreas},
  journal = {Rev. Mod. Phys.},
  volume = {93},
  issue = {2},
  pages = {025005},
  numpages = {72},
  year = {2021},
  month = {May},
  publisher = {American Physical Society},
  doi = {10.1103/RevModPhys.93.025005},
  url = {https://link.aps.org/doi/10.1103/RevModPhys.93.025005}
}

@article{fast_storage_cavities,
   title={Fast storage of photons in cavity-assisted quantum memories},
   volume={22},
   ISSN={2331-7019},
   url={http://dx.doi.org/10.1103/PhysRevApplied.22.044038},
   DOI={10.1103/physrevapplied.22.044038},
   number={4},
   journal={Physical Review Applied},
   publisher={American Physical Society (APS)},
   author={Kollath-Bönig, Johann S. and Dellantonio, Luca and Giannelli, Luigi and Schmit, Tom and Morigi, Giovanna and Sørensen, Anders S.},
   year={2024},
   month=oct }

@ARTICLE{qt_communication,
  author={Hasan, Syed Rakib and Chowdhury, Mostafa Zaman and Saiam, Md. and Jang, Yeong Min},
  journal={IEEE Access}, 
  title={Quantum Communication Systems: Vision, Protocols, Applications, and Challenges}, 
  year={2023},
  volume={11},
  number={},
  pages={15855-15877},
  doi={10.1109/ACCESS.2023.3244395}}

@article{Ringbauer2022,
archivePrefix = {arXiv},
arxivId = {2109.06903},
author = {Ringbauer, Martin and Meth, Michael and Postler, Lukas and Stricker, Roman and Blatt, Rainer and Schindler, Philipp and Monz, Thomas},
doi = {10.1038/s41567-022-01658-0},
eprint = {2109.06903},
issn = {17452481},
journal = {Nature Physics},
number = {9},
pages = {1053--1057},
title = {{A universal qudit quantum processor with trapped ions}},
volume = {18},
year = {2022}
}

@article{high_level_qc,
author = {Luo, Mingxing and Wang, Xiaojun},
doi = {10.1007/s11433-014-5551-9},
isbn = {1143301455519},
issn = {16747348},
journal = {Science China: Physics, Mechanics and Astronomy},
number = {9},
pages = {1712--1717},
title = {{Universal quantum computation with qudits}},
volume = {57},
year = {2014}
}

@article{Chen_2021,
	author = {Wentao Chen and Jaren Gan and Jing-Ning Zhang and Dzmitry Matuskevich and Kihwan Kim},
	doi = {10.1088/1674-1056/ac01e3},
	journal = {Chinese Physics B},
	month = {jun},
	number = {6},
	pages = {060311},
	publisher = {Chinese Physical Society and IOP Publishing Ltd},
	title = {Quantum computation and simulation with vibrational modes of trapped ions},
	url = {https://dx.doi.org/10.1088/1674-1056/ac01e3},
	volume = {30},
	year = {2021}}

@ARTICLE{no_control_infinite_dimensional_systems,
  author={Bloch, Anthony M. and Brockett, Roger W. and Rangan, Chitra},
  journal={IEEE Transactions on Automatic Control}, 
  title={Finite Controllability of Infinite-Dimensional Quantum Systems}, 
  year={2010},
  volume={55},
  number={8},
  pages={1797-1805},
  doi={10.1109/TAC.2010.2044273}}

@article{PhysRevLett.71.1816,
  title = {Quantum state engineering of the radiation field},
  author = {Vogel, K. and Akulin, V. M. and Schleich, W. P.},
  journal = {Phys. Rev. Lett.},
  volume = {71},
  issue = {12},
  pages = {1816--1819},
  numpages = {0},
  year = {1993},
  month = {Sep},
  publisher = {American Physical Society},
  doi = {10.1103/PhysRevLett.71.1816},
  url = {https://link.aps.org/doi/10.1103/PhysRevLett.71.1816}
}

@article{jc_controlabillity,
  title={Approximate controllability of the Jaynes-Cummings dynamics},
  author={Pinna, Lorenzo and Panati, Gianluca},
  journal={Journal of Mathematical Physics},
  volume={59},
  number={7},
  year={2018},
  publisher={AIP Publishing}
}

@article{PhysRevA.90.023824,
  title = {Arbitrary-quantum-state preparation of a harmonic oscillator via optimal control},
  author = {Rojan, Katharina and Reich, Daniel M. and Dotsenko, Igor and Raimond, Jean-Michel and Koch, Christiane P. and Morigi, Giovanna},
  journal = {Phys. Rev. A},
  volume = {90},
  issue = {2},
  pages = {023824},
  numpages = {11},
  year = {2014},
  month = {Aug},
  publisher = {American Physical Society},
  doi = {10.1103/PhysRevA.90.023824},
  url = {https://link.aps.org/doi/10.1103/PhysRevA.90.023824}
}

@article{qt_preparation_steering, title={State Preparation on Quantum Computers via Quantum Steering}, volume={5}, rights={https://creativecommons.org/licenses/by/4.0/legalcode}, ISSN={2689-1808}, DOI={10.1109/TQE.2024.3358193}, abstractNote={Quantum computers present a compelling platform for the study of open quantum systems, namely the non-unitary dynamics of a system. Here, we investigate and report digital simulations of Markovian, non-unitary dynamics that converge to a unique steady state. The steady state is programmed as a desired target state, yielding semblance to a quantum state preparation protocol. By delegating ancilla qubits and systems qubits, the system state is driven to the target state by repeatedly performing the following steps: (1) executing a designated system-ancilla entangling circuit, (2) measuring the ancilla qubits, and (3) re-initializing ancilla qubits to known states through active reset. While the ancilla qubits are measured and reinitialized to known states, the system qubits undergo a non-unitary evolution and are steered from arbitrary initial states to desired target states. We show results of the method by preparing arbitrary qubit states and qutrit (three-level) states on contemporary quantum computers. We also demonstrate that the state convergence can be accelerated by utilizing the readouts of the ancilla qubits to guide the protocol in a non-blind manner. Our work serves as a nontrivial example that incorporates and characterizes essential operations such as qubit reuse (qubit reset), entangling circuits, and measurement. These operations are not only vital for near-term noisy intermediate-scale quantum (NISQ) applications but are also crucial for realizing future error-correcting codes.}, journal={IEEE Transactions on Quantum Engineering}, author={Volya, Daniel and Mishra, Prabhat}, year={2024}, pages={1–14}}

@article{ivan_jc_hamiltonian, title={Invariant approach to the driven Jaynes-Cummings model}, volume={16}, ISSN={2542-4653}, DOI={10.21468/SciPostPhys.16.1.007}, abstractNote={We investigate the dynamics of the driven Jaynes-Cummings model, where a two-level atom interacts with a quantized field and both, atom and field, are driven by an external classical field. Via an invariant approach, we are able to transform the corresponding Hamiltonian into the one of the standard Jaynes-Cummings model. Subsequently, the exact analytical solution of the Schrödinger equation for the driven system is obtained and employed to analyze some of its dynamical variables.}, number={1}, journal={SciPost Physics}, author={Bocanegra-Garay, Ivan Alejandro and Hernández-Sánchez, L. and Ramos-Prieto, Irán and Soto-Eguibar, Francisco and Moya-Cessa, Héctor Manuel}, year={2024}, month=jan, pages={007}}

@article{PhysRevA.83.032302,
  title = {Quantum-state preparation with universal gate decompositions},
  author = {Plesch, Martin and Brukner, \ifmmode \check{C}\else \v{C}\fi{}aslav},
  journal = {Phys. Rev. A},
  volume = {83},
  issue = {3},
  pages = {032302},
  numpages = {5},
  year = {2011},
  month = {Mar},
  publisher = {American Physical Society},
  doi = {10.1103/PhysRevA.83.032302},
  url = {https://link.aps.org/doi/10.1103/PhysRevA.83.032302}
}

@article{Porotti2022deepreinforcement,
  doi = {10.22331/q-2022-06-28-747},
  url = {https://doi.org/10.22331/q-2022-06-28-747},
  title = {Deep {R}einforcement {L}earning for {Q}uantum {S}tate {P}reparation with {W}eak {N}onlinear {M}easurements},
  author = {Porotti, Riccardo and Essig, Antoine and Huard, Benjamin and Marquardt, Florian},
  journal = {{Quantum}},
  issn = {2521-327X},
  publisher = {{Verein zur F{\"{o}}rderung des Open Access Publizierens in den Quantenwissenschaften}},
  volume = {6},
  pages = {747},
  month = jun,
  year = {2022}
}

@article{PhysRevResearch.5.043002,
  title = {Sample-efficient model-based reinforcement learning for quantum control},
  author = {Khalid, Irtaza and Weidner, Carrie A. and Jonckheere, Edmond A. and Schirmer, Sophie G. and Langbein, Frank C.},
  journal = {Phys. Rev. Res.},
  volume = {5},
  issue = {4},
  pages = {043002},
  numpages = {21},
  year = {2023},
  month = {Oct},
  publisher = {American Physical Society},
  doi = {10.1103/PhysRevResearch.5.043002},
  url = {https://link.aps.org/doi/10.1103/PhysRevResearch.5.043002}
}

@article{PRXQuantum.2.010101,
  title = {From Pulses to Circuits and Back Again: A Quantum Optimal Control Perspective on Variational Quantum Algorithms},
  author = {Magann, Alicia B. and Arenz, Christian and Grace, Matthew D. and Ho, Tak-San and Kosut, Robert L. and McClean, Jarrod R. and Rabitz, Herschel A. and Sarovar, Mohan},
  journal = {PRX Quantum},
  volume = {2},
  issue = {1},
  pages = {010101},
  numpages = {16},
  year = {2021},
  month = {Jan},
  publisher = {American Physical Society},
  doi = {10.1103/PRXQuantum.2.010101},
  url = {https://link.aps.org/doi/10.1103/PRXQuantum.2.010101}
}

@article{PhysRevA.109.022441,
  title = {Universal composite pulses for robust quantum state engineering in four-level systems},
  author = {Shi, Zhi-Cheng and Wang, Jian-Hui and Zhang, Cheng and Song, Jie and Xia, Yan},
  journal = {Phys. Rev. A},
  volume = {109},
  issue = {2},
  pages = {022441},
  numpages = {15},
  year = {2024},
  month = {Feb},
  publisher = {American Physical Society},
  doi = {10.1103/PhysRevA.109.022441},
  url = {https://link.aps.org/doi/10.1103/PhysRevA.109.022441}
}

@article{PhysRevA.100.023410,
  title = {Composite pulses with errant phases},
  author = {Torosov, Boyan T. and Vitanov, Nikolay V.},
  journal = {Phys. Rev. A},
  volume = {100},
  issue = {2},
  pages = {023410},
  numpages = {9},
  year = {2019},
  month = {Aug},
  publisher = {American Physical Society},
  doi = {10.1103/PhysRevA.100.023410},
  url = {https://link.aps.org/doi/10.1103/PhysRevA.100.023410}
}

@software{jax2018github,
  author = {James Bradbury and Roy Frostig and Peter Hawkins and Matthew James Johnson and Chris Leary and Dougal Maclaurin and George Necula and Adam Paszke and Jake Vander{P}las and Skye Wanderman-{M}ilne and Qiao Zhang},
  title = {{JAX}: composable transformations of {P}ython+{N}um{P}y programs},
  url = {http://github.com/jax-ml/jax},
  version = {0.3.13},
  year = {2018},
}

@article{LEVITT198661,
	author = {Malcolm H. Levitt},
	doi = {https://doi.org/10.1016/0079-6565(86)80005-X},
	issn = {0079-6565},
	journal = {Progress in Nuclear Magnetic Resonance Spectroscopy},
	number = {2},
	pages = {61-122},
	title = {Composite pulses},
	url = {https://www.sciencedirect.com/science/article/pii/007965658680005X},
	volume = {18},
	year = {1986}}

@book{lie_algebras_sun, address={Basel}, title={The Lie Algebras su(N)}, rights={http://www.springer.com/tdm}, ISBN={978-3-7643-2418-6}, url={http://link.springer.com/10.1007/978-3-0348-8097-8}, DOI={10.1007/978-3-0348-8097-8}, publisher={Birkhäuser Basel}, author={Pfeifer, Walter}, year={2003}, pages={16,108}}

@misc{haar_measure,
      title={How to generate random matrices from the classical compact groups},
      author={Francesco Mezzadri},
      year={2007},
      eprint={math-ph/0609050},
      archivePrefix={arXiv},
      primaryClass={math-ph},
      url={https://arxiv.org/abs/math-ph/0609050},
}

@article{many_qudits_paper,
	author = {Vargas-Calder\'{o}n ,Vladimir and Parra-A. ,Nicolas and Vinck-Posada ,Herbert and Gonz\'{a}lez ,Fabio A.},
	doi = {10.7566/JPSJ.90.114002},
	eprint = {https://doi.org/10.7566/JPSJ.90.114002},
	journal = {Journal of the Physical Society of Japan},
	number = {11},
	pages = {114002},
	title = {Many-Qudit Representation for the Travelling Salesman Problem Optimisation},
	url = {https://doi.org/10.7566/JPSJ.90.114002},
	volume = {90},
	year = {2021}}

@article{PhysRevLett.76.1055,
  title = {Arbitrary Control of a Quantum Electromagnetic Field},
  author = {Law, C. K. and Eberly, J. H.},
  journal = {Phys. Rev. Lett.},
  volume = {76},
  issue = {7},
  pages = {1055--1058},
  numpages = {0},
  year = {1996},
  month = {Feb},
  publisher = {American Physical Society},
  doi = {10.1103/PhysRevLett.76.1055},
  url = {https://link.aps.org/doi/10.1103/PhysRevLett.76.1055}
}

@article{qt_metrology,
	author = {Couteau, Christophe and Barz, Stefanie and Durt, Thomas and Gerrits, Thomas and Huwer, Jan and Prevedel, Robert and Rarity, John and Shields, Andrew and Weihs, Gregor},
	date = {2023/06/01},
	doi = {10.1038/s42254-023-00589-w},
	id = {Couteau2023},
	isbn = {2522-5820},
	journal = {Nature Reviews Physics},
	number = {6},
	pages = {354--363},
	title = {Applications of single photons in quantum metrology, biology and the foundations of quantum physics},
	url = {https://doi.org/10.1038/s42254-023-00589-w},
	volume = {5},
	year = {2023}}

@book{garcía_ripoll_2022, place={Cambridge}, title={Quantum Information and Quantum Optics with Superconducting Circuits}, DOI={10.1017/9781316779460}, publisher={Cambridge University Press}, author={García Ripoll, Juan José}, year={2022}}

@article{QML-real-world,
	author = {Cerezo, M. and Verdon, Guillaume and Huang, Hsin-Yuan and Cincio, Lukasz and Coles, Patrick J.},
	da = {2022/09/01},
	doi = {10.1038/s43588-022-00311-3},
	id = {Cerezo2022},
	isbn = {2662-8457},
	journal = {Nature Computational Science},
	number = {9},
	pages = {567--576},
	title = {Challenges and opportunities in quantum machine learning},
	ty = {JOUR},
	url = {https://doi.org/10.1038/s43588-022-00311-3},
	volume = {2},
	year = {2022}}

@article{https://doi.org/10.1002/qute.201900038,
	author = {Cozzolino, Daniele and Da Lio, Beatrice and Bacco, Davide and Oxenl{\o}we, Leif Katsuo},
	doi = {https://doi.org/10.1002/qute.201900038},
	eprint = {https://onlinelibrary.wiley.com/doi/pdf/10.1002/qute.201900038},
	journal = {Advanced Quantum Technologies},
	number = {12},
	pages = {1900038},
	title = {High-Dimensional Quantum Communication: Benefits, Progress, and Future Challenges},
	url = {https://onlinelibrary.wiley.com/doi/abs/10.1002/qute.201900038},
	volume = {2},
	year = {2019}}

@article{quantum_chemistry,
  title = {Initial State Preparation for Quantum Chemistry on Quantum Computers},
  author = {Fomichev, Stepan and Hejazi, Kasra and Zini, Modjtaba Shokrian and Kiser, Matthew and Fraxanet, Joana and Casares, Pablo Antonio Moreno and Delgado, Alain and Huh, Joonsuk and Voigt, Arne-Christian and Mueller, Jonathan E. and Arrazola, Juan Miguel},
  journal = {PRX Quantum},
  volume = {5},
  issue = {4},
  pages = {040339},
  numpages = {37},
  year = {2024},
  month = {Dec},
  publisher = {American Physical Society},
  doi = {10.1103/PRXQuantum.5.040339},
  url = {https://link.aps.org/doi/10.1103/PRXQuantum.5.040339}
}

@misc{build_Quantum_supercomputer,
      title={How to Build a Quantum Supercomputer: Scaling Challenges and Opportunities}, 
      author={Masoud Mohseni and Artur Scherer and K. Grace Johnson and Oded Wertheim and Matthew Otten and Navid Anjum Aadit and Kirk M. Bresniker and Kerem Y. Camsari and Barbara Chapman and Soumitra Chatterjee and Gebremedhin A. Dagnew and Aniello Esposito and Farah Fahim and Marco Fiorentino and Abdullah Khalid and Xiangzhou Kong and Bohdan Kulchytskyy and Ruoyu Li and P. Aaron Lott and Igor L. Markov and Robert F. McDermott and Giacomo Pedretti and Archit Gajjar and Allyson Silva and John Sorebo and Panagiotis Spentzouris and Ziv Steiner and Boyan Torosov and Davide Venturelli and Robert J. Visser and Zak Webb and Xin Zhan and Yonatan Cohen and Pooya Ronagh and Alan Ho and Raymond G. Beausoleil and John M. Martinis},
      year={2024},
      eprint={2411.10406},
      archivePrefix={arXiv},
      primaryClass={quant-ph},
      url={https://arxiv.org/abs/2411.10406}, 
}

@article{Quantum_Simulators,
  title = {Quantum Simulators: Architectures and Opportunities},
  author = {Altman, Ehud and Brown, Kenneth R. and Carleo, Giuseppe and Carr, Lincoln D. and Demler, Eugene and Chin, Cheng and DeMarco, Brian and Economou, Sophia E. and Eriksson, Mark A. and Fu, Kai-Mei C. and Greiner, Markus and Hazzard, Kaden R.A. and Hulet, Randall G. and Koll\'ar, Alicia J. and Lev, Benjamin L. and Lukin, Mikhail D. and Ma, Ruichao and Mi, Xiao and Misra, Shashank and Monroe, Christopher and Murch, Kater and Nazario, Zaira and Ni, Kang-Kuen and Potter, Andrew C. and Roushan, Pedram and Saffman, Mark and Schleier-Smith, Monika and Siddiqi, Irfan and Simmonds, Raymond and Singh, Meenakshi and Spielman, I.B. and Temme, Kristan and Weiss, David S. and Vu\ifmmode \check{c}\else \v{c}\fi{}kovi\ifmmode \acute{c}\else \'{c}\fi{}, Jelena and Vuleti\ifmmode \acute{c}\else \'{c}\fi{}, Vladan and Ye, Jun and Zwierlein, Martin},
  journal = {PRX Quantum},
  volume = {2},
  issue = {1},
  pages = {017003},
  numpages = {19},
  year = {2021},
  month = {Feb},
  publisher = {American Physical Society},
  doi = {10.1103/PRXQuantum.2.017003},
  url = {https://link.aps.org/doi/10.1103/PRXQuantum.2.017003}
}

@article{PRXQuantum.4.027001,
  title = {Quantum Simulation for High-Energy Physics},
  author = {Bauer, Christian W. and Davoudi, Zohreh and Balantekin, A. Baha and Bhattacharya, Tanmoy and Carena, Marcela and de Jong, Wibe A. and Draper, Patrick and El-Khadra, Aida and Gemelke, Nate and Hanada, Masanori and Kharzeev, Dmitri and Lamm, Henry and Li, Ying-Ying and Liu, Junyu and Lukin, Mikhail and Meurice, Yannick and Monroe, Christopher and Nachman, Benjamin and Pagano, Guido and Preskill, John and Rinaldi, Enrico and Roggero, Alessandro and Santiago, David I. and Savage, Martin J. and Siddiqi, Irfan and Siopsis, George and Van Zanten, David and Wiebe, Nathan and Yamauchi, Yukari and Yeter-Aydeniz, K\"ubra and Zorzetti, Silvia},
  journal = {PRX Quantum},
  volume = {4},
  issue = {2},
  pages = {027001},
  numpages = {70},
  year = {2023},
  month = {May},
  publisher = {American Physical Society},
  doi = {10.1103/PRXQuantum.4.027001},
  url = {https://link.aps.org/doi/10.1103/PRXQuantum.4.027001}
}

@article{PhysRevA.110.012430,
  title = {Quantum algorithm for solving the advection equation using Hamiltonian simulation},
  author = {Brearley, Peter and Laizet, Sylvain},
  journal = {Phys. Rev. A},
  volume = {110},
  issue = {1},
  pages = {012430},
  numpages = {12},
  year = {2024},
  month = {Jul},
  publisher = {American Physical Society},
  doi = {10.1103/PhysRevA.110.012430},
  url = {https://link.aps.org/doi/10.1103/PhysRevA.110.012430}
}

@article{Krovi2023improvedquantum,
  doi = {10.22331/q-2023-02-02-913},
  url = {https://doi.org/10.22331/q-2023-02-02-913},
  title = {Improved quantum algorithms for linear and nonlinear differential equations},
  author = {Krovi, Hari},
  journal = {{Quantum}},
  issn = {2521-327X},
  publisher = {{Verein zur F{\"{o}}rderung des Open Access Publizierens in den Quantenwissenschaften}},
  volume = {7},
  pages = {913},
  month = feb,
  year = {2023}
}

@article{PhysRevLett.127.090504,
  title = {High-Efficiency Arbitrary Quantum Operation on a High-Dimensional Quantum System},
  author = {Cai, W. and Han, J. and Hu, L. and Ma, Y. and Mu, X. and Wang, W. and Xu, Y. and Hua, Z. and Wang, H. and Song, Y. P. and Zhang, J.-N. and Zou, C.-L. and Sun, L.},
  journal = {Phys. Rev. Lett.},
  volume = {127},
  issue = {9},
  pages = {090504},
  numpages = {7},
  year = {2021},
  month = {Aug},
  publisher = {American Physical Society},
  doi = {10.1103/PhysRevLett.127.090504},
  url = {https://link.aps.org/doi/10.1103/PhysRevLett.127.090504}
}

@article{PhysRevA.109.042401,
  title = {Efficient quantum state preparation with Walsh series},
  author = {Zylberman, Julien and Debbasch, Fabrice},
  journal = {Phys. Rev. A},
  volume = {109},
  issue = {4},
  pages = {042401},
  numpages = {22},
  year = {2024},
  month = {Apr},
  publisher = {American Physical Society},
  doi = {10.1103/PhysRevA.109.042401},
  url = {https://link.aps.org/doi/10.1103/PhysRevA.109.042401}
}

@article{stp_matrix_products,
	author = {Iaconis, Jason and Johri, Sonika and Zhu, Elton Yechao},
	date = {2024/01/25},
	doi = {10.1038/s41534-024-00805-0},
	id = {Iaconis2024},
	isbn = {2056-6387},
	journal = {npj Quantum Information},
	number = {1},
	pages = {15},
	title = {Quantum state preparation of normal distributions using matrix product states},
	url = {https://doi.org/10.1038/s41534-024-00805-0},
	volume = {10},
	year = {2024}}

@article{qt_chesmestry_2,
	author = {Bauer, Bela and Bravyi, Sergey and Motta, Mario and Chan, Garnet Kin-Lic},
	date = {2020/11/25},
	doi = {10.1021/acs.chemrev.9b00829},
	isbn = {0009-2665},
	journal = {Chemical Reviews},
	journal1 = {Chemical Reviews},
	journal2 = {Chem. Rev.},
	month = {11},
	number = {22},
	pages = {12685--12717},
	publisher = {American Chemical Society},
	title = {Quantum Algorithms for Quantum Chemistry and Quantum Materials Science},
	type = {doi: 10.1021/acs.chemrev.9b00829},
	url = {https://doi.org/10.1021/acs.chemrev.9b00829},
	volume = {120},
	year = {2020},
	year1 = {2020}}

@article{vlado_classification_measurments,
	author = {Gonz{\'a}lez ,Fabio A. and Vargas-Calder{\'o}n ,Vladimir and Vinck-Posada ,Herbert},
	date = {2021/04/15},
	doi = {10.7566/JPSJ.90.044002},
	isbn = {0031-9015},
	journal = {Journal of the Physical Society of Japan},
	journal1 = {Journal of the Physical Society of Japan},
	journal2 = {J. Phys. Soc. Jpn.},
	month = {2025/01/11},
	n2 = {This paper reports a novel method for supervised machine learning based on the mathematical formalism that supports quantum mechanics. The method uses projective quantum measurement as a way of building a prediction function. Specifically, the relationship between input and output variables is represented as the state of a bipartite quantum system. The state is estimated from training samples through an averaging process that produces a density matrix. Prediction of the label for a new sample is made by performing a projective measurement on the bipartite system with an operator, prepared from the new input sample, and applying a partial trace to obtain the state of the subsystem representing the output. The method can be seen as a generalization of Bayesian inference classification and as a type of kernel-based learning method. One remarkable characteristic of the method is that it does not require learning any parameters through optimization. We illustrate the method with different 2-D classification benchmark problems and different quantum information encodings.},
	number = {4},
	pages = {044002},
	publisher = {The Physical Society of Japan},
	title = {Classification with Quantum Measurements},
	type = {doi: 10.7566/JPSJ.90.044002},
	url = {https://doi.org/10.7566/JPSJ.90.044002},
	volume = {90},
	year = {2021},
	year1 = {2021}}

@inbook{Shore_2011,
	author = {Shore, Bruce W.},
	booktitle = {Manipulating Quantum Structures Using Laser Pulses},
	pages = {97--136},
	place = {Cambridge},
	publisher = {Cambridge University Press},
	title = {Two-state coherent excitation},
	year = {2011}}

@article{neural_network_training,
	author = {Narkhede, Meenal V. and Bartakke, Prashant P. and Sutaone, Mukul S.},
	da = {2022/01/01},
	doi = {10.1007/s10462-021-10033-z},
	id = {Narkhede2022},
	isbn = {1573-7462},
	journal = {Artificial Intelligence Review},
	number = {1},
	pages = {291--322},
	title = {A review on weight initialization strategies for neural networks},
	ty = {JOUR},
	url = {https://doi.org/10.1007/s10462-021-10033-z},
	volume = {55},
	year = {2022}}

@misc{zhou2024optimalcontrolopenquantum,
      title={Optimal Control for Open Quantum System in Circuit Quantum Electrodynamics}, 
      author={Mo Zhou and F. A. Cárdenas-López and Sugny Dominique and Xi Chen},
      year={2024},
      eprint={2412.20149},
      archivePrefix={arXiv},
      primaryClass={quant-ph},
      url={https://arxiv.org/abs/2412.20149}, 
}

@article{PhysRevLett.122.020502,
  title = {Black-Box Quantum State Preparation without Arithmetic},
  author = {Sanders, Yuval R. and Low, Guang Hao and Scherer, Artur and Berry, Dominic W.},
  journal = {Phys. Rev. Lett.},
  volume = {122},
  issue = {2},
  pages = {020502},
  numpages = {5},
  year = {2019},
  month = {Jan},
  publisher = {American Physical Society},
  doi = {10.1103/PhysRevLett.122.020502},
  url = {https://link.aps.org/doi/10.1103/PhysRevLett.122.020502}
}

@article{Bausch2022fastblackboxquantum,
  doi = {10.22331/q-2022-08-04-773},
  url = {https://doi.org/10.22331/q-2022-08-04-773},
  title = {Fast {B}lack-{B}ox {Q}uantum {S}tate {P}reparation},
  author = {Bausch, Johannes},
  journal = {{Quantum}},
  issn = {2521-327X},
  publisher = {{Verein zur F{\"{o}}rderung des Open Access Publizierens in den Quantenwissenschaften}},
  volume = {6},
  pages = {773},
  month = aug,
  year = {2022}
}

@article{sparse_preparation,
	author = {de Veras, Tiago M. L. and da Silva, Leon D. and da Silva, Adenilton J.},
	date = {2022/06/14},
	doi = {10.1007/s11128-022-03549-y},
	id = {de Veras2022},
	isbn = {1573-1332},
	journal = {Quantum Information Processing},
	number = {6},
	pages = {204},
	title = {Double sparse quantum state preparation},
	url = {https://doi.org/10.1007/s11128-022-03549-y},
	volume = {21},
	year = {2022}

}

@article{PhysRevA.110.032609,
  title = {Simple quantum algorithm to efficiently prepare sparse states},
  author = {Ramacciotti, Debora and Lefterovici, Andreea I. and Rotundo, Antonio F.},
  journal = {Phys. Rev. A},
  volume = {110},
  issue = {3},
  pages = {032609},
  numpages = {10},
  year = {2024},
  month = {Sep},
  publisher = {American Physical Society},
  doi = {10.1103/PhysRevA.110.032609},
  url = {https://link.aps.org/doi/10.1103/PhysRevA.110.032609}
}

@misc{goswami2025quditbasedscalablequantumalgorithm,
      title={Qudit-based scalable quantum algorithm for solving the integer programming problem}, 
      author={Kapil Goswami and Peter Schmelcher and Rick Mukherjee},
      year={2025},
      eprint={2508.13906},
      archivePrefix={arXiv},
      primaryClass={quant-ph},
      url={https://arxiv.org/abs/2508.13906}, 
}

@misc{wang2025machinelearningassistedpulsedesignstate,
      title={ }, 
      author={Zhao-Wei Wang and Hong-Yang Ma and Yun-An Yan and Lian-Ao Wu and Zhao-Ming Wang},
      year={2025},
      eprint={2508.20377},
      archivePrefix={arXiv},
      primaryClass={quant-ph},
      url={https://arxiv.org/abs/2508.20377}, 
}

@article{universalapprox,
  title={Multilayer feedforward networks are universal approximators},
  author={Hornik, Kurt and Stinchcombe, Maxwell and White, Halbert},
  journal={Neural Networks},
  volume={2},
  number={5},
  pages={359--366},
  year={1989},
  doi={10.1016/0893-6080(89)90020-8},
  url={https://doi.org/10.1016/0893-6080(89)90020-8}
}

\onecolumn
\appendix
\section{Neural network}
\label{app:nns}
In this work, we consider a feed-forward neural network, which results from the consecutive application of neural network layers.
A neural network layer is a parameterized function $l:\mathbb{R}^\text{in}\to\mathbb{R}^\text{out}$ defined as
\begin{align}
  l(\vb*{x}) = a(W\vb*{x} + \vb{b}),
\end{align}
where $a$ is an element-wise function, called the activation function, $W\in\mathbb{R}^\text{out}\times\mathbb{R}^\text{in}$ is called the weight matrix, and $\vb*{b}\in\mathbb{R}^\text{out}$ is the bias vector.
Both the weight matrix and the bias vector contain the parameters of the layer.
Layers can be composed to form a feed-forward neural network, which is the type considered in this work.
Such a neural network is defined as
\begin{align}
  f_{\vb*{\eta}}(\vb*{x}) = l_L\circ l_{L-1}\circ\cdots\circ l_1(\vb*{x})
\end{align}
for a neural network of $L$ layers.
We have made it explicit that $\vb*{\eta}$ are the neural network's parameters, which are all the weight matrices and bias vectors from the neural network's layers.
The last layer's output size is $3N$, to accommodate the $3N$ pulse parameters $\vb*{\Theta}=\{\phi_i\}_{i=1}^N\cup\{\varphi_i\}_{i=1}^N\cup\{t_i\}_{i=1}^N$.
The first layer's input is $d$, where $d$ refers to the components that specify the input target state.
The neural-network architecture depends on the target dimension $n$ and was found by optimizing the network structure with Optuna. For $n=2$ we used hidden-layer widths $[209, 173, 222, 347]$; for $n=3$ we used $[197, 264, 304]$; and for $n=4$ we used $[321, 442, 425]$.

\section{Basis of SU($n$)}
\label{app:sun}
The SU($n$) basis that we consider
is the union of the following $n\times n$ families of matrices  \cite{lie_algebras_sun}:

\begin{equation}
  \begin{pmatrix}
    0 & 1 & \cdots & 0 & 0\\
    1 & 0 & \cdots & 0 & 0\\
    \vdots & \vdots & \ddots & \vdots & \vdots\\
    0 & 0 & \cdots & 0 & 0\\
    0 & 0 & \cdots & 0 & 0
  \end{pmatrix},
  \begin{pmatrix}
    0 & 0 & 1 & \cdots & 0\\
    0 & 0 & 0 & \cdots & 0\\
    1 & 0 & 0 & \cdots & 0\\
    0 & \vdots & \ddots & \vdots & \vdots\\
    0 & 0 & \cdots & 0 & 0
  \end{pmatrix},
  \cdots
  \begin{pmatrix}
    0 & 0 & \cdots & 0 & 0\\
    0 & 0 & \cdots & 0 & 0\\
    \vdots & \vdots & \ddots & \vdots & \vdots\\
    0 & 0 & \cdots & 0 & 1\\
    0 & 0 & \cdots & 1 & 0
  \end{pmatrix}
\end{equation}

\begin{equation}
  \begin{pmatrix}
    0 & -i & \cdots & 0 & 0\\
    i & 0 & \cdots & 0 & 0\\
    \vdots & \vdots & \ddots & \vdots & \vdots\\
    0 & 0 & \cdots & 0 & 0\\
    0 & 0 & \cdots & 0 & 0
  \end{pmatrix},
  \begin{pmatrix}
    0 & 0 & -i & \cdots & 0\\
    0 & 0 & 0 & \cdots & 0\\
    i & 0 & 0 & \cdots & 0\\
    0 & \vdots & \ddots & \vdots & \vdots\\
    0 & 0 & \cdots & 0 & 0
  \end{pmatrix},
  \cdots
  \begin{pmatrix}
    0 & 0 & \cdots & 0 & 0\\
    0 & 0 & \cdots & 0 & 0\\
    \vdots & \vdots & \ddots & \vdots & \vdots\\
    0 & 0 & \cdots & 0 & -i\\
    0 & 0 & \cdots & i & 0
  \end{pmatrix}
\end{equation}

\begin{equation}
  \begin{pmatrix}
    1 & 0 & 0 & \cdots & 0 \\
    0 & -1 & 0 &\cdots & 0 \\
    0 & 0 & 0 &\cdots & 0 \\
    \vdots & \vdots & \vdots&  \ddots & 0 \\
    0 & 0 &  0 & \cdots & 0 \\
  \end{pmatrix},
  \sqrt{\frac{1}{3}}
  \begin{pmatrix}
    1 & 0 & 0 & \cdots & 0 \\
    0 & 1 & 0 &\cdots & 0 \\
    0 & 0 & -2 &\cdots & 0 \\
    \vdots & \vdots & \vdots&  \ddots & 0 \\
    0 & 0 &  0 & \cdots & 0 \\
  \end{pmatrix}
  ,
  \cdots,
  \sqrt{\frac{2}{n(n-1)}}
  \begin{pmatrix}
    1 & 0 & 0 & \cdots & 0 \\
    0 & 1 & 0 &\cdots & 0 \\
    0 & 0 & 1 &\cdots & 0 \\
    \vdots & \vdots & \vdots&  \ddots & \vdots \\
    0 & 0 &  0 & \cdots & -n+1 \\
  \end{pmatrix}
\end{equation}

This basis corresponds to $n^2-1$ matrices. \added{Using this basis, for a target state $\ket{\psi_{\text{target}}}$, the inputs for the neural network are the $n^2-1$ expectation values $\expval{G_k}{\psi_{\text{target}}}$, where $\{G_k\}_{k=1}^{n^2-1}$ denotes the SU($n$) basis listed above}.

\end{document}